\documentclass[journal]{vgtc}                     

\usepackage{mathptmx}                  
\usepackage{graphicx}                  
\usepackage{times}                     

\usepackage{amsmath,amsfonts,amssymb}
\usepackage{algorithmic}
\usepackage{algorithm}
\usepackage{array}
\usepackage{textcomp}
\usepackage{stfloats}
\usepackage{url}
\usepackage{verbatim}
\usepackage{lipsum}
\usepackage{booktabs}
\usepackage{tabu}
\usepackage{tabularx}
\usepackage{listings}
\usepackage[dvipsnames]{xcolor} 
\usepackage{longtable}
\usepackage{tikz}
\usepackage{colortbl}
\usepackage{multirow}
\usepackage{fontawesome6} 
\usepackage{makecell}
\usepackage{ltablex}   
\keepXColumns          
\usepackage{tcolorbox}
\usepackage{subcaption} 

\graphicspath{{figs/}{figures/}{pictures/}{images/}{./}} 

\usepackage[subtle]{savetrees}

\newcommand{\provega}{\textsc{ProVega}}
\newcommand{\req}[1]{$\mathbb{R}$#1}
\newcommand{\cat}[1]{\textit{\bfseries#1}}
\newcommand{\code}[1]{\texttt{\small\textcolor{black}{#1}}}
\definecolor{taxonomyColorProgression}{HTML}{377eb8}
\definecolor{taxonomyColorVisualization}{HTML}{e41a1c}
\definecolor{taxonomyColorInteraction}{HTML}{4daf4a}
\definecolor{taxonomyColorGuidance}{HTML}{ff7f00}

\newcommand{\mypar}[1]{\vspace{0mm}\noindent\textbf{{#1}}\hspace{1mm}}

\newtcbox{\codepill}{
  on line,
  boxsep=1pt,
  left=3pt,
  right=3pt,
  top=1pt,
  bottom=0pt,
  colback=gray!10,
  colframe=gray!40,
  arc=3pt,
  boxrule=0.4pt,
  fontupper=\ttfamily\small
}

\newcommand{\pdiv}[4]{%
    \begin{tikzpicture}[baseline=2pt, xscale=0.4, yscale=0.4]
        \colorlet{pdivColor}{black!70}
        
        \fill[pdivColor] (0,0) rectangle (1,#1);
        \draw (0,0) rectangle (1,1);
        \node[text=pdivColor] at (0.5, -0.6) {\tiny \textbf{P}};
        
        \fill[pdivColor] (1.2,0) rectangle (2.2,#2);
        \draw (1.2,0) rectangle (2.2,1);
        \node[text=pdivColor] at (1.7, -0.6) {\tiny \textbf{D}};
        
        \fill[pdivColor] (2.4,0) rectangle (3.4,#3);
        \draw (2.4,0) rectangle (3.4,1);
        \node[text=pdivColor] at (2.9, -0.6) {\tiny \textbf{I}};
        
        \fill[pdivColor] (3.6,0) rectangle (4.6,#4);
        \draw (3.6,0) rectangle (4.6,1);
        \node[text=pdivColor] at (4.1, -0.6) {\tiny \textbf{V}};
    \end{tikzpicture}%
}

\title{ProVega: A Grammar to Ease the Prototyping, Creation, and Reproducibility of Progressive Data Analysis and Visualization Solutions}

\author{
  \authororcid{Matteo Filosa$^\mathbf{*}$}{0009-0004-8868-7907}, 
  \authororcid{Graziano Blasilli$^\mathbf{*}$}{0000-0003-3339-6403}, 
  \authororcid{Emilio Martino}{0009-0006-9009-2501}, 
  and \authororcid{Marco Angelini}{0000-0001-9051-6972}
}

\authorfooter{
\item[\textbf{*}] Matteo Filosa and Graziano Blasilli are both first authors.\vspace{2mm}
\item Filosa, Blasilli, and Martino are with Sapienza University of Rome, Italy. 
\item Angelini is with Link Campus University, Rome, Italy.
}

\abstract{%
Modern data analysis requires speed for massive datasets. Progressive Data Analysis and Visualization (PDAV) emerged as a discipline to address this problem, providing fast response times while maintaining interactivity with controlled accuracy. Yet it remains difficult to implement and reproduce. To lower this barrier, we present \provega, a Vega-Lite-based grammar that simplifies PDAV instrumentation for both simple visualizations and complex visual environments. Alongside it, we introduce Pro-Ex, an editor designed to streamline the creation and analysis of progressive solutions. We validated \provega\ by reimplementing 11 exemplars from the literature—verified for fidelity by 39 users—and demonstrating its support for various progressive methods, including data-chunking, process-chunking, and mixed-chunking. An expert user study confirmed the efficacy of \provega\ and the Pro-Ex environment in real-world tasks. \provega, Pro-Ex, and all related materials are available at \url{https://github.com/XAIber-lab/provega}.
}


\renewcommand{\manuscriptnotetxt}{}

\begin{document}



\maketitle


\section{Introduction}
\label{sec:introduction}

Data analysis always faces the challenge of processing increasingly large datasets while simultaneously meeting stringent response-time requirements for both analysis and visualization~\cite{6842585,10003102}. To bridge this gap, Progressive Data Analysis and Visualization (PDAV) (aka Progressive Visual Analytics)~\cite{Stolper2014,DBLP:journals/corr/FeketeP16,Angelini2018Review} has emerged as a specialized discipline that, by leveraging partial results, prioritizes rapid feedback and fluid interactivity while controlling and communicating accuracy. This field recently reached a milestone of maturity with the publication of a comprehensive book~\cite{Fekete2024} that guides researchers and practitioners through the entire design process, from conceptual foundations to rigorous evaluation.

Despite these advantages, PDAV is hindered by the high level of technical expertise required to build even basic or prototypical solutions. This inherent complexity makes it difficult to reproduce state-of-the-art research when an implementation is unavailable or to adapt existing tools for progressiveness to test its potential benefits before committing to it, ultimately stalling its adoption. To mitigate these issues, this paper introduces \provega, a grammar based on Vega-Lite that simplifies implementing progressive workflows.

\provega\ is built on a systematic collection of requirements drawn from the existing literature and uses a flexible architecture of internal properties and hooks. This design allows users to craft solutions ranging from simple, progressive visualizations using default settings to more sophisticated ones that integrate custom analytical components, orchestrated through the grammar. 
To further support users in crafting progressive solutions with \provega, we also contribute Pro-Ex, an explorer and editor that streamlines the creation, analysis, and sharing of PDAV solutions. This assisted environment helps quickly retrieve existing progressive implementations as a basis for reuse and modification, or to craft the solution from scratch if none of the available ones are satisfactory. Additionally, it serves as an onboarding environment for novice users, providing support across the various implementation stages with \provega\ and enabling the saving of work history and intermediate states for didactic purposes, such as teaching or training users on PDAV.

We validated \provega’s capability for prototyping and reproducibility by re-implementing 11 exemplars from PDAV literature. This effort included recreating systems that lacked publicly available source code, thereby demonstrating how to implement previously unavailable solutions in a reproducible form. These systems were then evaluated by 39 users to ensure they maintained high fidelity to the original works. Additionally, we present four use cases to demonstrate how \provega\ supports the primary methods of progression: data-chunking, process-chunking, the novel mixed-chunking introduced by Ulmer et al. \cite{Ulmer2024}, and integration with external frameworks and tools when custom solutions are required. Finally, an expert study spanning multiple sessions and tasking users to create a real implementation confirmed the efficacy of \provega\ in addressing progressive design needs, as well as the effectiveness and usability of the Pro-Ex environment with real data and tasks. 

\noindent
Summarizing, this paper contributes:
\begin{itemize}
    \item \provega, a Vega-lite compatible grammar to manage the prototyping and implementation of PDAV solutions, from simple to complex, completely integrated or linked to external dedicated modules, covering all requirements of a PDAV solution through a systematic collection.
    \item Pro-Ex, an interactive environment to retrieve, explore, and implement PDAV solutions using \provega, providing support to novice and expert users, and supporting implementation, sharing, and onboarding tasks;
    \item Evaluation of \provega\ and Pro-Ex, through four use cases and two user evaluations, showing efficacy, effectiveness, and usability of the provided tools.
\end{itemize}

\noindent
All tools and materials, including \provega\ and Pro-Ex, are available at \url{https://github.com/XAIber-lab/provega}.

\section{Related Work}

\provega\ sits at the intersection of three research areas: the theory and design of progressive data analysis and visualization, the declarative grammars for creating visualizations, and the assisted environments that leverage progressive computation.

\mypar{Progressive Data Analysis and Visualization}
Progressive Data Analysis and Visualization (PDAV) emerged as a paradigm for reconciling the scale of modern datasets with the interactive demands of exploratory analysis \cite{Stolper2014,DBLP:journals/corr/FeketeP16,Angelini2018Review}.
PDA splits long computations into a series of partial results that improve over time, so that analysts receive early feedback and can act or steer before the full computation completes.
Unlike eager systems that block until completion and streaming systems that react to continuously arriving data, a progressive function delivers increasingly accurate intermediate outputs on a static dataset, supports steering (e.g., parameter changes) between steps without triggering a full restart, and enables early decisions and continuous hypothesis testing on massive data.
Coupled with visualization, PDA becomes \emph{Progressive Visual Analytics} (PVA)~\cite{Stolper2014,8943144}.
Schulz et al. \cite{Schulz2016} extended the design by distinguishing \emph{data chunking} from \emph{process chunking}, and Ulmer et al. \cite{Ulmer2024} introduced \emph{mixed chunking} as a third strategy.
Angelini et al. \cite{Angelini2018Review} synthesized the earlier PVA landscape into a set of 45 design requirements spanning data, processing, visualization, and interaction, which directly ground the \provega\ grammar.
This field recently reached a milestone of maturity with the publication of a comprehensive survey of progressive visualizations by Ulmer et al. \cite{Ulmer2024}, and a comprehensive book~\cite{Fekete2024} that guides researchers and practitioners through the entire design process, from conceptual foundations to rigorous evaluation.

\mypar{Grammars}
Declarative grammars have become a central paradigm for creating data visualizations, leveraging concise descriptions of visualizations through composable abstractions rather than imperative code. 
The foundations of modern visualization grammars trace back to Wilkinson’s \textit{Grammar of Graphics}~\cite{Wilkinson1999}, which introduced a formal model for constructing statistical graphics through a set of combinatorial components, including data transformations, graphical marks, and visual encoding channels. 
Several solutions subsequently exploited Wilkinson’s model. Polaris \cite{Stolte2002}, later commercialized as Tableau, introduced a grammar-based interface for constructing table-based visualization of relational data.
In parallel, ggplot2~\cite{ggplot2}, a new data visualization package for R that uses the insights from Wilkinson's Grammar of Graphics, has emerged.
More recent work has extended grammar-based approaches to support interactive and web-based visualization systems.
Satyanarayan et al. proposed Vega~\cite{Satyanarayan2016}, a declarative grammar for interactive graphics based on a reactive dataflow model, enabling dynamic visualizations driven by event streams and data transformations. Vega provides a low-level but expressive specification language for defining visualization, interaction logic, and rendering behavior.
Building on Vega, Vega-Lite~\cite{VegaLite} provides a higher-level grammar that simplifies the specification of interactive visualizations in JSON. This design allows for concise specifications while supporting a broad range of interaction techniques and visualizations.
Subsequent work further extended the grammar to support dynamic visual behaviors. 
Animated Vega-Lite~\cite{Zong2023AnimatedVegaLite} extended the Vega-Lite grammar to create animated visualizations. The approach models animation as time-varying data queries and represents time either as an encoding channel or as an event stream that drives selections and transformations. 
Despite these advances, existing visualization grammars primarily focus on specifying static or interactive visualizations but do not explicitly address progressive data analysis and visualization workflows. 

\mypar{Assisted Environments}
Assisted environments in visualization enable building visual solutions, providing helpers to the user to leverage techniques~\cite{informatics6010014} to design for interactive exploration of large datasets, addressing latency challenges in exploratory analysis~\cite{lazarik2025,10.1145/3769841}.
For example, recent AI-assisted frameworks like VizCV enable end-to-end analysis of publication trajectories by leveraging topic evolution, impact metrics, and collaboration networks, with automated insights generated via LLMs~\cite{lazarik2025}.
Immersive, no-code tools, such as VisualSphere for clinical data, further exemplify assisted systems that recommend visualizations without programming expertise~\cite{visualsphere2024}. In this landscape, Pro-Ex plays a complementary role, allowing users to explore progressive visualizations from the literature, familiarize themselves with the progressive grammar, and then implement their own progressive visualizations with ease (\cref{sec:age}).
These solutions are tailored to specific domains and do not generalize.
In contrast, grammar-based environments are domain-independent. While grammar-based solutions provide their editors (e.g., Vega-Editor), these environments usually support only basic editing and are not necessarily tailored toward progressive workflows.
The PDAV field presents a limited number of assisted environments: ProgressiVis by Fekete~\cite{progressivis}, proposed in 2015 and recently updated (2025), provides a Python-based environment to support the development of progressive workflows. Unlike our approach, it is not based on a grammar and still requires programming skills to be exploited. It is also a full-code environment, but it still does not provide facilities for exploring and reusing existing solutions, as Pro-Ex does. 
We additionally show in \cref{sec:usecases} how \provega\ and Pro-Ex can be coupled with general-purpose progressive backends, such as ProgressiVi, to drive progressive workflows.
%
Li and Ma introduced P4~\cite{Li2020P4}, a web-based toolkit that pairs a declarative JSON grammar with GPU computing to accelerate both data transformation and rendering. P5~\cite{Li2020P5} extends it with support to progressive parallel processing: the source data is partitioned into chunks processed and rendered incrementally, so partial results are available immediately, and analysts can steer or stop the computation at any time.
Both systems share with \provega\ the use of a declarative grammar, but pursue different primary goals. P4 and P5 are performance-oriented toolkits whose contribution is a GPU-accelerated engine exposed through a JavaScript API; \provega\ is a declarative grammar that describes the full PDAV workflow, and optionally delegates computation to pluggable backends.


\section{The \provega\ Grammar Design}
\label{sec:design}

The design of the \provega\ grammar is based on a systematic collection of requirements for \textit{Progressive Data Analysis and Visualization} (PDAV). 
%
We build upon the set of 45 requirements originally introduced by Angelini et al.~\cite{Angelini2018Review} for Progressive Visual Analytics (PVA) systems.
We then reviewed the recent literature and identified 19 additional requirements, expanding the scope from PVA to PDAV. Together, these 64 requirements form the basis of a hierarchical taxonomy that captures the conceptual space of PDAV and informs the design of the \provega\ grammar. 

\subsection{PDAV Requirements: Extending from PVA}
\label{sec:pva_req}

Angelini et al.~\cite{Angelini2018Review} introduced a set of 45 requirements for Progressive Visual Analytics (PVA), synthesizing key goals and design criteria. Although originally scoped for visual analytics systems, many of these requirements apply to broader progressive workflows involving data computation, transformation, and user interaction. 
We mapped them to the topics of the Progressive Data Analysis book~\cite{Fekete2024} to verify their applicability to PDAV (see supplemental materials).
Therefore, we adopt these as the foundational layer of our PDAV requirement set.
For their in-depth discussion, we refer the reader to their original definition~\cite{Angelini2018Review}, while we report here the complete list along with their source: 
\req{Hel1} \cite{Hellerstein1999}, \req{H2}–\req{H6} \cite{Hetzler2005}, \req{C7}–\req{C8} \cite{Chandramouli2013}, \req{F9}–\req{F12} \cite{Ferreira2014}, \req{S13}–\req{S20} \cite{Stolper2014}, \req{M21}–\req{M31} \cite{Mühlbacher2014}, \req{T32}–\req{T41} \cite{Turkay2017}, \req{B42}–\req{B45} \cite{Badam2017}. 
%

These requirements are numbered sequentially (1-45), and grouped in four broad categories: \textit{data}, \textit{processing}, \textit{visualization}, and \textit{interaction}. 
Some requirements belong exclusively to one category, while others belong to two, reflecting the integrated nature of PVA solutions.

\textit{Data} requirements encompass aspects of data ingestion and subdivision, as well as prioritization and aggregation strategies, in a progressive solution.
\textit{Processing} requirements encompass all aspects, from the progressive implementation of computation to its execution and control.
\textit{Visualization} requirements encompass all aspects, from visual feedback on the running process to the dynamic presentation of incremental outcomes.
\textit{Interaction} requirements encompass all aspects from meeting human time constraints to providing structured interactions with the process.

To capture the latest developments in the PDAV field, we extended the original list of requirements \cite{Angelini2018Review} based on a literature analysis, using the recently released Progressive Data Analysis book~\cite{Fekete2024} and recent publications on progresiveness that target its design principles as primary sources. The goal of this extended review was to enrich the theoretical foundation upon which we aim to build our grammar.
We identified 19 additional requirements, bringing the total to 64.
These new requirements focus on three previously underexplored areas: quality indicators throughout the progression, support for different progressive data processing, and user guidance mechanisms.
In the appendix, \cref{tab:app:new_pdva_req} shows details about the new requirements. For their in-depth discussion, we refer the reader to the original publications and limit this section to a high-level overview. 

We followed the nomenclature proposed by Angelini et al.~\cite{Angelini2018Review}, continuing to number them sequentially:
\begin{itemize*}
    \item \req{A46}-\req{A49}: Angelini et al., 2019 \cite{Angelini2019_OnQuality}
    \item \req{Sch50}-\req{Sch51}: Schulz et al., 2016 \cite{Schulz2016Enhanced}
    \item \req{UL52}: Ulmer et al., 2024 \cite{Ulmer2024}
    \item \req{PM53}-\req{PM59}: Pérez-Messina et al., 2024 \cite{PerezMessina2024Enhancing}
    \item \req{PM60}-\req{PM64}: Pérez-Messina et al., 2025 \cite{PerezMessina2025Coupling} 
\end{itemize*}

Requirements \req{A46}-\req{A49} highlight the role of quality indicators during the progression. These requirements emphasize the importance of informing users about the ongoing computation, specifically regarding its progress, stability, and the certainty of partial results.

A second group of requirements targets the chunking strategies used to implement progression. Three distinct chunking types could exist: \textit{data chunking}(\req{Sch50}), where subsets of the data are incrementally loaded and processed; \textit{process chunking} (\req{Sch51}), where the whole dataset is involved in successive computational steps; and \textit{mixed chunking} (\req{UL52}), which combines both strategies. 

Finally, a third set of requirements emphasizes the role of guidance in progressive data analysis and visualization by sustaining and supporting users in maintaining flow when solving complex, data-intensive analytical tasks.
These requirements distinguish between the two guidance approaches \cite{PerezMessina2025Coupling}: \textit{Guidance for Progressiveness} (G4P), where guidance enhances and enriches user tasks during progression \req{PM53}-\req{PM59}, and \textit{Progressiveness for Guidance} (P4G), where the progression itself supports the generation of guidance \req{PM60}-\req{PM64}. 
Each category is further articulated through the three guidance degrees \cite{Ceneda2017}: \textit{orienting} (provides cues through an additional information layer), \textit{directing} (suggests where to focus next), and \textit{prescribing} (recommends specific actions).

\subsection{ProVega Requirements Taxonomy}
\label{sec:taxonomy}

The 64 collected requirements, which propose both functional needs and design principles, served as the basis for the grammar development process.
We organized them into a three-level hierarchical taxonomy, grouping them into conceptually related categories and subcategories. 
This taxonomy aims to capture the conceptual landscape of PDAV and defines the structure of the grammar specification, supporting modularity and extensibility and serving as a framework for adding future features and new grammar properties.
Additionally, it is organized to reflect a designer's point of view, making it easier to identify the characteristics she wants implemented in her progressive solution.

\begin{figure}[t]
    \vspace{-2mm}
    \centering
    \includegraphics[width=0.9\linewidth]{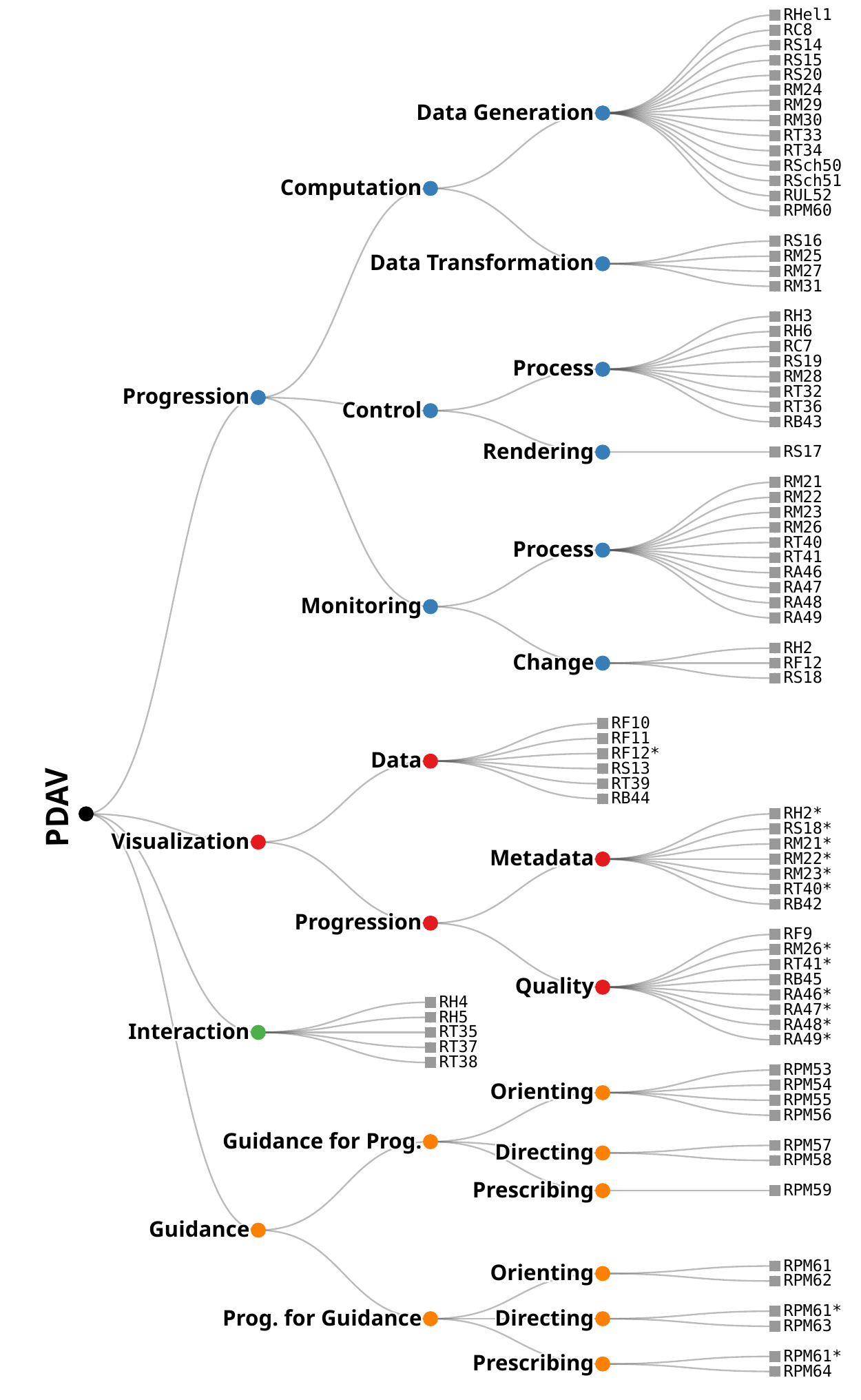}
    \vspace{-2mm}
    \caption{
PDAV requirements taxonomy, serving as the foundation of the \provega\ grammar definition. PDAV requirements (\cref{sec:pva_req}) are structured into four categories: {\color{taxonomyColorProgression}Progression}, {\color{taxonomyColorVisualization}Visualization}, {\color{taxonomyColorInteraction}Interaction}, and {\color{taxonomyColorGuidance}Guidance}.
Each is further subdivided into subcategories across three hierarchical levels.
Bigger figure in the appendix (\cref{fig:app:taxonomy_big}).
}
\label{fig:taxonomy}
\end{figure}

To design the taxonomy, we began by reorganizing the requirements into four main categories: \cat{Progression}, \cat{Visualization}, \cat{Interaction}, and \cat{Guidance}. Within each of these primary categories, we further refined the structure by identifying meaningful subcategories, where applicable, to better capture the functional diversity of the requirements. We iteratively repeated the hierarchical process, resulting in the three-level taxonomy shown in \cref{fig:taxonomy}. 
While most requirements belong to a single principal category, 13 span two categories: \cat{Progression} and \cat{Visualization}. To better preserve the mapping with the original requirements formulation, we duplicated them. On the contrary, the implemented part is specific to the subcategory they belong to. In \cref{fig:taxonomy}, duplicated requirements are marked with an asterisk (*) to distinguish them from their original occurrence.

\mypar{\color{taxonomyColorProgression}Progression}
This category involves 39 requirements related to the data \cat{Computation}, as well as the \cat{Control} and \cat{Monitoring} of the progressive process. This category focuses on how the system incrementally produces results, how to steer the process, and how users are informed about the process. 
Within the \cat{Computation} category, we distinguish between two computational phases: \cat{Data Generation} and \cat{Data Transformation}. They differentiate between producing intermediate results and subsequently refining or manipulating those results before presenting them visually.
The \cat{Data Generation} subcategory involves requirements that rule how data is incrementally produced, prioritized, and steered during progression. This includes support for the different chunking strategies: data chunking (\req{Sch50}), process chunking (\req{Sch51}), and mixed chunking (\req{UL52}). Moreover, this subcategory includes requirements aimed at enhancing early usefulness, such as prioritizing interesting data (\req{Hel1}, \req{M29}) or allowing subspace selection (\req{S15}, \req{S20}).
The \cat{Data Transformation} subcategory involves transforming and refining the partial results produced by mechanisms such as subspace filtering (\req{S16}) and data aggregation (\req{M25}).
The \cat{Control} category focuses on enabling the steering of the progressive process.
We distinguish between two subcategories: \cat{Process Control} and \cat{Rendering Control}, where the first concerns controlling the execution of the progressive computation, and the second concerns controlling the rendering of its outcomes.
\cat{Process Control} includes a set of requirements that allow users to influence the timing, content, and execution of the progression. For instance, requirements such as \req{H3} and \req{H6} ask for giving users explicit control over the arrival of updates and the ability to dynamically adjust them, while \req{S19} supports an on-demand refresh strategy, allowing analysts to choose when they are ready to see the latest results while the progressive process continues behind the scene. Other requirements, such as \req{M28}, introduce mechanisms for execution cancellation, which are essential for avoiding unnecessary computation. 
Furthermore, progressive solutions should also support multiple operational modes, such as monitoring and exploration (\req{B43}).
In contrast, \cat{Rendering Control} addresses how visual updates are handled. \req{S17}, for example, explains the need to minimize distractions by avoiding excessive view changes.
The \cat{Monitoring} category encompasses requirements that enable the user to observe and understand the state and evolution of the progressive process without directly steering it. 
Unlike the \cat{Control} category, which relies on user intervention, \cat{Monitoring} focuses on passive awareness and feedback mechanisms that enable users to monitor execution status, interpret uncertainty, and track changes in the visualization. 
We divide \cat{Monitoring} into two subcategories: \cat{Process Monitoring} and \cat{Change Monitoring}.
The \cat{Process Monitoring} includes requirements that provide information about the execution’s status and quality. These involves aspects such as process aliveness (\req{M21}, \req{T40}), absolute and relative progress (\req{M22}, \req{M23}, \req{A46}, \req{A47}), and uncertainty associated with the results (\req{M26}, \req{T41}, \req{A48}, \req{A49}). 
In contrast, the \cat{Change Monitoring} subcategory involves how updates in the visualization are communicated. For instance, \req{H2} and \req{S18} underline mechanisms for detecting and highlighting what is new or has changed in the visualization after each update.
For all these subcategories, the focus is on governing the data and process aspects, neglecting the visualization.

\mypar{\color{taxonomyColorVisualization}Visualization}
This category includes 21 requirements that address how visual representations support the progressive process. 
These requirements involve both representing the underlying data and metadata, such as quality information, and communicating progress information.
The category is divided into two subcategories: \cat{Data Visualization} and \cat{Progression Visualization}.
The \cat{Data Visualization} subcategory focuses on how the underlying data is visually represented throughout the analysis, ensuring continuity, clarity, and interpretability as results evolve. Several key requirements define the design principles for this area. 
For instance, \req{F10} accentuates the importance of consistency in visualizations across different tasks; \req{F11} requires maintaining the spatial stability of visualizations; \req{F12} recommends minimizing visual noise; \req{S13} states the importance of aligning visual updates with the user’s cognitive workflow.
The \cat{Progression Visualization} subcategory contains requirements that underscore the role of visual elements in communicating the internal dynamics of progressive computation and informing the user about the state, stability, and quality of the progressive process itself. We divide this area into two subcategories: \cat{Progression Metadata} and \cat{Progression Quality}.
The \cat{Progression Metadata Visualization} subcategory includes visual feedback elements that help users track the computation's evolution over time. For instance, \req{H2} requires that users can easily see what changes or updates have occurred with each new increment of processed data. At the same time, \req{S18} suggests providing visual cues to highlight newly generated results. Requirements such as \req{M21}, \req{M22}, and \req{M23} specify the need for feedback mechanisms that indicate whether the computation is running and its progress state. 
The \cat{Progression Quality Visualization} subcategory focuses on the reliability, stability, and uncertainty associated with partial results. Requirements such as \req{F9}, \req{M26}, and \req{T41} express the importance of communicating uncertainty. \req{A46}-\req{A49} emphasize the importance of visualizing indicators of progress, stability, and certainty.
%

\mypar{\color{taxonomyColorInteraction}Interaction}
This category includes five requirements that emphasize the need for fluid user interaction throughout the progressive process. 
Unlike other categories, \cat{Interaction} is not further subdivided, as prescriptions from literature are still limited.
For instance, \req{H4} requests minimizing disruption to the analytic process and interaction flows; \req{T37} recommends designing interactions that account for fluctuations; and \req{T38} suggests providing interaction mechanisms to define a structured investigation sequence. 

\mypar{\color{taxonomyColorGuidance}Guidance}
This category comprises 13 PDAV requirements that specifically address the role of guidance. These requirements are divided into two subcategories: \cat{Guidance for Progressiveness (G4P)}  and \cat{Progressiveness for Guidance (P4G)}.
Each of these subcategories is further structured according to three degrees of guidance: \cat{Orienting}, \cat{Directing}, and \cat{Prescribing}.

\section{The ProVega Grammar}
\label{sec:grammar}

The \provega\ grammar formalizes the conceptual space of Progressive Data Analysis and Visualization (PDAV) as identified through the taxonomy presented in \cref{sec:taxonomy}.
It defines a set of properties that enable the declarative specification of progressive behaviors built on the Vega-Lite \cite{VegaLite} grammar. 
For each requirement in the taxonomy, we evaluated whether a grammar property could be defined.
The relationship between requirements and grammar properties is not always one-to-one. In certain cases, a single grammar property addresses multiple requirements, indicating that diverse needs may converge on a shared formal construct.
Conversely, some requirements do not result in the definition of a specific grammar property. This situation arises when a requirement falls outside the \provega's scope or can be addressed by existing Vega-Lite constructs without further extensions. 
Therefore, the grammar represents a higher-level formalization rather than a direct transcription of the complete set of requirements. 
A detailed summary of requirement coverage is provided as supplemental material.
%
A detailed summary of requirement coverage and complete documentation of each grammar property, its options, and examples are available as supplemental material.

\subsection{Grammar Structure}
\label{sec:grammar-structure}

\provega\ extends the JSON syntax of Vega-Lite specifications (\textbf{specs}) by adding custom properties.
To preserve compatibility and minimize interference with Vega-Lite specifications, all progressive-related constructs are encapsulated under a single top-level property: \texttt{provega}. This approach allows the core Vega-Lite semantics to remain intact and helps progression features to be selectively enabled or disabled without altering the underlying Vega-Lite visualization logic.
Each top-level category of the taxonomy (\cat{\color{taxonomyColorProgression}Progression}, \cat{\color{taxonomyColorVisualization}Visualization}, \cat{\color{taxonomyColorInteraction}Interaction}, and \cat{\color{taxonomyColorGuidance}Guidance}) is implemented as a corresponding sub-block within the \code{provega} object: \code{provega.progression}, \code{provega.visualization}, \code{provega.interaction}, and \code{provega.guidance}. 
This modular and hierarchical organization mirrors the structure of the conceptual taxonomy. It allows properties to be grouped semantically and supports future extensibility. In the following, we describe the grammar properties defined under each main category.

To improve readability and avoid repetition, we adopt a shorthand notation throughout the remainder of the paper. When referring to a grammar property using the textual style format \codepill{X}, we implicitly include the \code{provega} namespace: thus, the property \codepill{X} stands for \codepill{provega.X}. 

\mypar{\color{taxonomyColorProgression}Progression}
The \codepill{progression} block defines how data is processed incrementally, how progression can be steered or controlled, and how users are informed of the execution status. This block corresponds to the \cat{Progression} category of the taxonomy.
%
In Vega-Lite, the \code{data} property specifies the input data, which can be either inline (\code{values}) or remote (\code{url}). The \provega\ grammar extends this mechanism by supporting WebSocket URLs as data sources. This allows data to be streamed progressively from an external generator implementing any custom processing pipeline. We distinguish between two data-feeding modes: a \textit{complete input}, in which the entire dataset is embedded or retrieved at once, and a \textit{progressive input}, in which data items are received incrementally via a WebSocket connection.

By addressing \cat{Data Computation} requirements, the property \codepill{progression.chunking.type} specifies the chunking strategy, i.e., how the data is incrementally processed. It supports three modes (\textit{data} \req{Sch50}, \textit{process} \req{Sch51}, and \textit{mixed} \req{UL52}).
In scenarios where progressive data input is used (e.g., via WebSocket), the user may only define the chunking type; the external generator is responsible for producing and dispatching the data.
In contrast, when using a complete dataset (e.g., via \code{data.values}), the \provega\ engine handles chunking internally. In this case, the chunking type must be set to \code{data}, and additional properties can be defined to control how the dataset is consumed. The \codepill{progression.chunking.reading} allows to choose the sampling strategy (e.g., \textit{ascending}, \textit{descending}, \textit{random}), the desired chunk size, and the reading frequency (\req{C8}).

Requirements in \cat{Data Transformation} request support for manipulating the partial results.
In this context, Vega-Lite’s native \code{transform} operators are sufficient to support many requirements (\req{S16}, \req{M25}, \req{M27}). Since these functionalities are already covered by the Vega-Lite grammar, no additional \provega\ properties are introduced for data transformation.

The \cat{Control} subcategory encompasses user-driven mechanisms to steer the progression process. These are specified through the \codepill{progression.control} block, which includes properties to enable pausing (\req{H3}, \req{H6}, \req{S19}), stopping (\req{M28}), and defining the frequency of updates (\req{T32}). 
In cases where data is streamed from an external process (e.g., via WebSocket), \provega\ offers a feedback mechanism that allows the visualization to communicate with the generator by sending acknowledgments (ACKs). These ACKs can signal when the process can proceed to the next data chunk, thereby enabling flow control. This mechanism must be supported by the external process logic to be effective. The pause and stop functionalities are implemented as buttons rendered below the visualization.
When the whole data is embedded in the spec, progression is managed entirely by the \provega\ engine.
In scenarios where the generator frequency cannot be controlled, users can configure the rendering frequency independently using \codepill{progression.control.min\_rendering\_frequency} (\req{S17}).

The \cat{Monitoring} subcategory focuses on passive feedback mechanisms that help users understand the status and dynamics of the progressive computation.
The property \codepill{progression.monitoring} enables the specification of feedback components related to the execution state and quality of the progression process. Several sub-properties are available to configure which regard aliveness (\req{M21}), expected remaining iterations and time (\req{M22}, \req{T40}). For example, the user can
enable a horizontal progress bar inserted below the visualization to indicate the absolute progress of the computation (\req{M22}).
In addition, the \codepill{progression.monitoring.quality} block supports the communication of multiple quality indicators (\req{T41}) about progress (\req{A46}, \req{A47}), stability (\req{A48}), and certainty (\req{A49}).
Each of these measures can be bound to a variable name, i.e., a field present in the data.
When provided, \provega\ renders these quality indicators as small multiple line charts positioned on the right side of the visualization. This allows users to track the evolution of each metric over time, offering a compact view of the process behavior.
\provega\ also supports the detection and visual encoding of changes in the dataset or marks. 
The \codepill{progression.monitoring.change.mark} allows highlighting individual marks that are newly added or updated since the last rendering. This visual feedback is particularly useful, for example, in plots such as scatterplots (where new points appear) or bar charts (where bar lengths change), and addresses requirement \req{H2}.
The \codepill{progression.monitoring.change.area} allows highlighting the affected region of the visualization using a temporary colored overlay. This mechanism collects user attention to areas where significant changes have occurred (\req{S18}).

\mypar{\color{taxonomyColorVisualization}Visualization}
The \cat{Visualization} category of the taxonomy focuses on how progressive results (data, metadata, and information) are visually represented.
Several requirements in this category are related to the \cat{Progression} category, particularly within its \cat{Monitoring} subcategory. 
In fact, most of the information from the monitoring task is presented to the user as small ancillary elements alongside the main visualization.

For example, requirements such as \req{M21}, \req{M22}, \req{M23}, \req{T40}, and \req{T41} are fulfilled by properties under \codepill{progression.monitoring}, which provide visual indications of process aliveness, progress, estimated completion time, and uncertainty. 
Similarly, \req{H2} and \req{S18} are addressed by \codepill{progression.monitoring.change}, which highlights newly inserted or updated marks in the visualization and overlays areas where changes occur. 
In addition, \codepill{progression.monitoring.quality} allows users to visually represent quality indicators such as progress, stability, and certainty (\req{A46}–\req{A49}).

The property \codepill{visualization.visual\_stability} allows users to explicitly request spatial stability across updates \req{F11}, enabling the \provega\ engine to preserve the layout of visual marks as new data chunks are progressively incorporated, and minimizing disruptive changes. With this feature, \provega\ helps users maintain their mental model of the evolving dataset, which is particularly valuable in progressive visualizations where partial results are frequently updated.

Finally, some requirements such as \req{F9} (\textit{uncertainty visualizations should be easy to interpret}) and \req{F10} (\textit{visualizations should be consistent across tasks}) are satisfied indirectly. While \provega\ does not provide ad hoc properties or solutions, it provides the necessary hooks (e.g., visual channels) to implement effective visualizations using the existing Vega-Lite properties.

\mypar{\color{taxonomyColorInteraction}Interaction}
The \provega\ grammar does not introduce any new interaction mechanisms beyond those already available in Vega-Lite. It inherits Vega-Lite’s built-in support for user interaction, satisfying \req{RH4} and \req{RH5}.
Three interaction-related requirements, however, fall outside the scope of a progressive visualization grammar. 
Although no interaction-specific properties are currently implemented, the \codepill{interaction} block has been reserved to accommodate future extensions. This provides a dedicated place to integrate progressive interaction mechanisms as new features emerge, ensuring consistency with \provega's modular, extensible design.


\mypar{\color{taxonomyColorGuidance}Guidance}
At the current stage of development, the \provega\ grammar includes basic support, derived from classic guidance, for the \codepill{guidance} block, as the literature has not produced specific solutions for progressive guidance or guidance for progressiveness. However, this block has been explicitly reserved to support future extensions. 

\subsection{Implementation}
\label{sec:provega-implementation}

\provega\ is implemented at the Vega-embed layer to intercept data and specifications without modifying the Vega-Lite compiler or Vega runtime. During initialization, data is buffered, and the visualization begins empty, providing full control over data release while preserving the standard rendering pipeline. An async handler, \code{progressiveLoadingHandler}, normalizes grammar properties and injects monitoring signals into the view. The core \code{progressiveLoading} loop then manages chunking, streams rows through \code{vega.changeset().insert}, and updates additional properties (e.g., the progress/estimated time of completion (ETC) indicators or quality feedback signals). This architecture supports local data and WebSockets through a unified scheduling layer, allowing users to specify chunking types via \codepill{progression.chunking.type}. By hooking into the embedding layer, \provega\ grafts progressive functionality onto stock Vega-embed while keeping downstream runtimes intact.
\section{Progressive Examples made through ProVega} 
\label{sec:gallery}

To demonstrate the expressive power and practical reproducibility capabilities of \provega, we built a curated, complexity-aware gallery designed to challenge the grammar across multiple dimensions using key exemplars from the PDAV literature. The goal of this gallery is to verify that the grammar supports the range of progressive behaviors documented in the literature through a unified, declarative specification. 
Table~\cref{tab:visualization_pdiv} shows the visualizations, their originating paper(s), their types (from Ulmer et al.~\cite{Ulmer2024}), the requirements coverage in four fields (P = Processing, D = Data, I = Interaction, V = Visualization), and their complexity. Such a score was calculated by taking into account the visualization's intrinsic complexity, including the number of visual components, encodings, and marks used, their data representation, and the difficulty of handling the dataset to achieve progressivity in a real-time scenario. 
For instance, the \emph{Density Map}~\cite{Schulz2016} achieves PDIV coverage (1, 0.75, 1, 1) thanks to its full pipeline UI with quality metric plotted over time (P=1), incremental data processing (D=0.75), sliders/buttons for control (I=1), and rearrangeable views with quality feedback (V=1). Conversely, the \emph{Pie Chart}~\cite{Sindol2012} scores lower (0.5, 0.5, 0.25, 0.5) as it handles large-category increments progressively (P/D=0.5) but lacks advanced interaction or change cues (I/V low).




\begin{figure*}[t]
    \vspace{-2mm}
    \centering
    \includegraphics[width=0.85\linewidth]{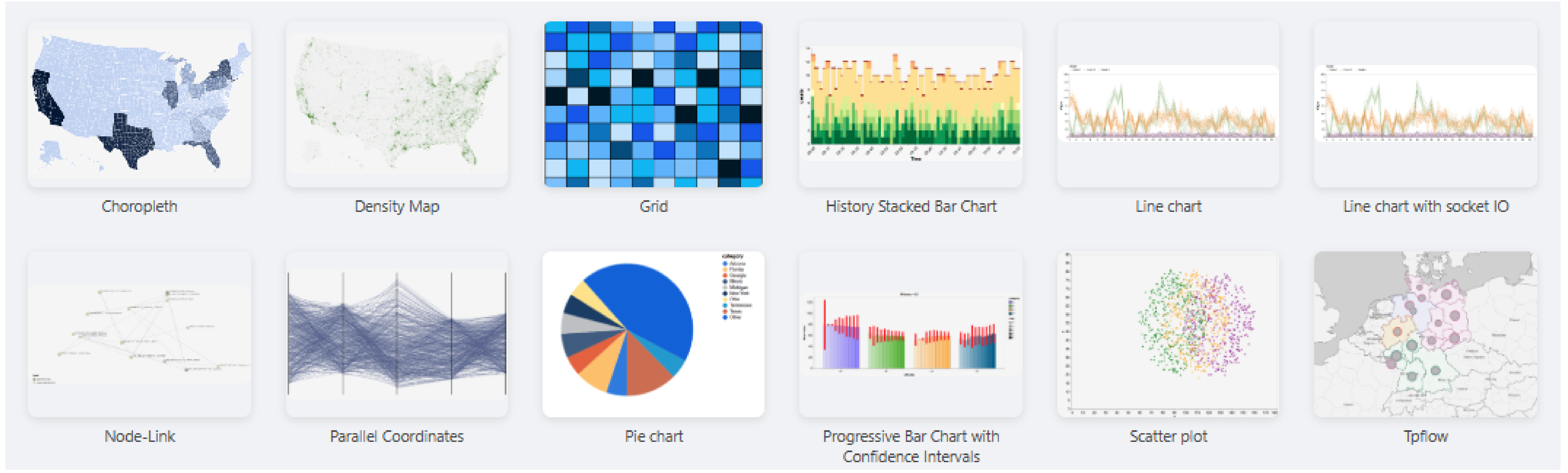}
    \vspace{-2mm}
    \caption{The Gallery View from the Pro-Ex, showing the 11 reimplemented exemplars from the PDAV literature} 
    \label{fig:browserView}
\end{figure*}

We picked a set of visualizations explicitly designed to span: (i) the main families of progressive techniques (data chunking, process chunking, mixed chunking), e.g., IncVisage Grid~\cite{Rahman2017} exemplifies data chunking with trendline increments; (ii) a complete coverage of progressive visualization types according to Ulmer et al.~\cite{Ulmer2024}; 
(iii) diverse data and task characteristics, from categorical aggregation (History Stacked Bar) to continuous distributions (Scatter Plot), from spatial/temporal data (Bubble Chart) to multidimensional ones (Line Chart); 
(iv) different progressive goals, such as preliminary utility (Density Map for early insights), consistency maintenance (Choropleth's iterative model), uncertainty communication (Scatter Plot's change coloring), and real-time monitoring (Node-Link's DOI expansion). These exemplars are explorable through the Pro-Ex environment.


\section{Use cases}
\label{sec:usecases}


We demonstrate the \provega\ grammar's capabilities across four use cases, aiming to cover a wide range of scenarios: data chunking, process chunking, mixed chunking, and integration with an external backend.

\subsection{Use Case 1: Data Chunking}

The \provega\ data chunking streams large datasets in small batches, while all other progressive behaviors, such as rendering, controls for progression, including play, pause, step forward, and backward, and the process quality signals, are governed by the top-level progressive block. For the data chunking use case, we employed a density map, leveraging the Fatality Analysis Reporting System (FARS)~\cite{nhtsa_fars_2025} dataset. It regards fatal injuries suffered in motor vehicle traffic crashes in America. The visualization in this use case is explorable in the Inspector View of Pro-Ex (see Section~\ref{sec:age}), and is one of the 11 reimplemented exemplars from the PDAV literature~\cite{Schulz2016}. In this use case, streaming the dataset in small batches enables the visualization to reveal spatial hotspots early (and refine them over time) even on heavy inputs: the used FARS dataset, reporting data from 2001 to 2009, contains on the order of $5.6\times 10^5$ rows, making a monolithic load-and-render noticeably slower to reach a first meaningful view. The no-hook pattern ingests raw data chunks without any user-defined callbacks, preserving a fully declarative workflow. 
Within this framework, the grammar defines several properties that govern the progressive flow: \code{provega.progression.chunking.type} enables row-based chunk insertion, while \code{provega.progression.chunking.reading.method} determines whether the system reads data sequentially or randomly. The parameters \code{provega.progression.chunking.reading.chunk\_size} and \code{provega.progression.chunking.reading.frequency} specifies, respectively, the number of rows per batch and the temporal interval between successive reads. 
The \code{provega.progression.control} property manages controls that steer chunk progression. Additional mechanisms are handled by \code{provega.progression.monitoring}, which conveys aliveness indicators, such as a spinner showing the estimated time to completion. Finally, \code{provega.progression.monitoring.change} controls zone or mark highlighting as each data chunk arrives, and \code{provega.progression.monitoring.quality} provides stage-wise signals describing absolute and relative progress, stability, and certainty throughout the execution.
At the end of the progression, the full visualization is available. Chunks can be manually added or removed using the step back and step forward buttons (provided by the \code{provega.progression.control} subproperties) to investigate, with fine-grained precision, how the insertion (or deletion) of a chunk affects the entire visualization. Our benchmarks demonstrate that \provega\ maintains a consistent visual update every 250ms, ensuring a steady and responsive refinement of the density map.


\begin{figure*}[!h]
    \vspace{-2mm}
    \centering
    \includegraphics[width=0.8\linewidth]{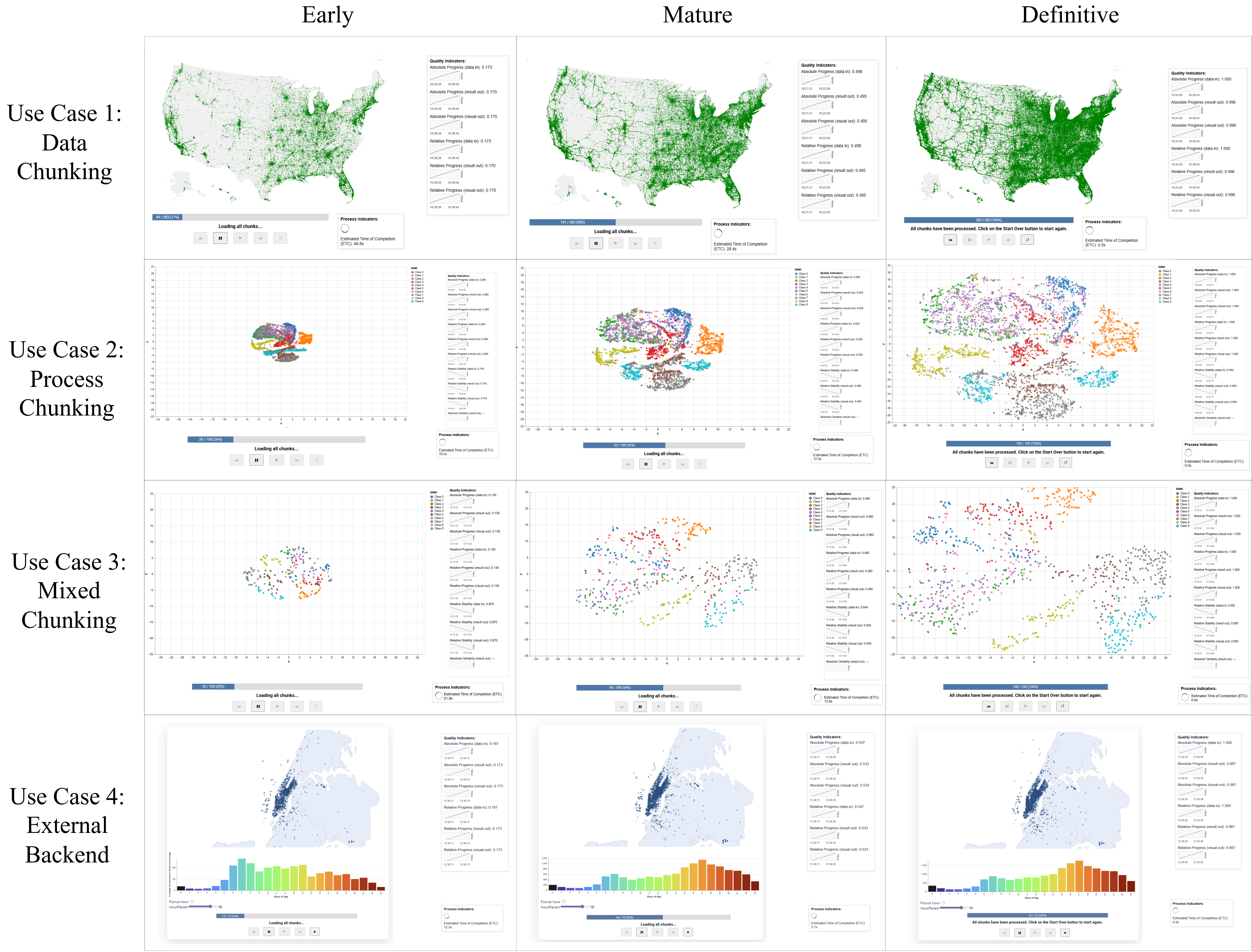}
    \vspace{-2mm}
    \caption{\provega\ rendering early, intermediate, and final states of progressive visualizations in the four use cases.
    }
    \label{fig:example}
\end{figure*}


\subsection{Use Case 2: Process Chunking}

We also implemented an example for the process chunking category, based on the t-distributed Stochastic Neighbor Embedding (t-SNE) clustering visualization~\cite{vandermaaten2008visualizing}, using the Fashion MNIST dataset~\cite{xiao2017fashion}. 
It is a large collection of grayscale images depicting 70,000 fashion items across 10 categories, such as shirts, shoes, and bags. 
The grammar has been shown to successfully replicate t-SNE on the Fashion MNIST dataset, visually exploring how different categories of fashion items cluster in a reduced-dimensional space.
\provega\ hook is used to link a JavaScript library, tsnejs ~\cite{karpathy_tsnejs} to compute a two-dimensional embedding, using standard parameter settings and a data-driven perplexity heuristic that adapts to dataset size. Computation is performed incrementally on all the data. After each iteration, the partial embedding is streamed to the Vega view, allowing progressive updates while keeping the interface responsive. 
Experimental results show that \provega\ achieves a rapid update cadence of 125ms, providing near-instantaneous visual feedback during the iterative optimization.



\subsection{Use Case 3: Mixed Chunking}

This use case builds on the previous one, but implements mixed chunking, explicitly combining data and process chunking within a single progressive workflow. Incoming or streamed records are incrementally integrated into the runtime dataset (data chunking), while the optimization itself advances in iterations (process chunking). After each batch, the current 2-D embedding is streamed to the Vega view, allowing the layout to evolve smoothly as both new data and optimization progress are incorporated. A stable mapping between records and embedding coordinates is maintained, ensuring that updates remain consistent as the dataset grows. The result is a progressive execution of the t-SNE clustering process in which both data arrival and optimization proceed incrementally, enabling continuous visual updates while avoiding costly full recomputations. Despite the complexity of the dual-stream process, the system successfully delivers a visual update every 500ms, keeping the evolving layout synchronized and interactive.

\subsection{Use Case 4: Progressive Data Engine Backend}

We tested \provega's capabilities to be coupled with an external backend implemented through ProgressiVis~\cite{progressivis}. It incrementally builds a \code{PTable}, a progressive, columnar table abstraction suited for progression, and streams plain JSON chunks over Socket.IO, with cadence and chunk size controlled by backend parameters (\code{CHUNK\_SIZE}, \code{CHUNK\_DELAY}, etc.) mapped in the corresponding \provega\ grammar properties. \provega\ intercepts data sources coming from Socket.IO, buffers chunks in an auxiliary buffer, and injects them into the named Vega dataset, while exposing UI controls (play/pause/back), process controls, and other additional properties provided by the grammar. In Figure~\ref{fig:example}, a map showing Uber pickups in April 2014 in New York City is progressively built, leveraging a public dataset~\cite{fivethirtyeight_uber_2014}. A bar chart shows their distribution by pickup hour; ticking the ``Focus hour'' checkbox will filter the pickups on the map by hour, which can be adjusted with the relative slider. 
The ProgressiVis backend loads the taxi CSV, computes derived fields (hour/weekday), and emits chunked records (\code{id, lat, lon, pickup\_ts, hour, weekday, batch}) over Socket.IO. \provega\ opens the socket, buffers incoming rows, and progressively commits them to the pickups dataset. Since ProgressiVis has no knowledge of marks or encodings, Vega-Lite progressively renders marks as the dataset grows. 
Any backend that can emit chunked JSON over WebSocket or Socket.IO can feed \provega\, and any Vega/Vega-Lite spec can be made progressive by adding \provega\ without changing marks or encodings. 
The integration sustains a reliable update interval of approximately 330ms, effectively translating backend data streams into smooth, incremental visual updates.
This demonstrates that \provega\ is general-purpose and can be integrated with any external backend to receive and control custom implementations of data/process/mixed chunking, as well as related analytics.


\section{The Pro-Ex Environment}
\label{sec:age}
In this Section, we introduce \provega's assisted grammar explorer, Pro-Ex, showing its three main views and functionalities.

\mypar{The Gallery View} The gallery view (see Figure~\ref{fig:browserView}) 
serves as the primary onboarding interface of the Pro-Ex, designed to explore a collection of available progressive exemplars. It provides filters for the visualization's name and category (e.g., hierarchical, layered, etc.), as well as the ability to upload new exemplars to the gallery. Each exemplar in the gallery can be loaded into the Inspector View to check its specification and be used as is or for further modification.

\begin{figure*}
    \vspace{-2mm}
    \centering
    \includegraphics[width=0.95\linewidth]{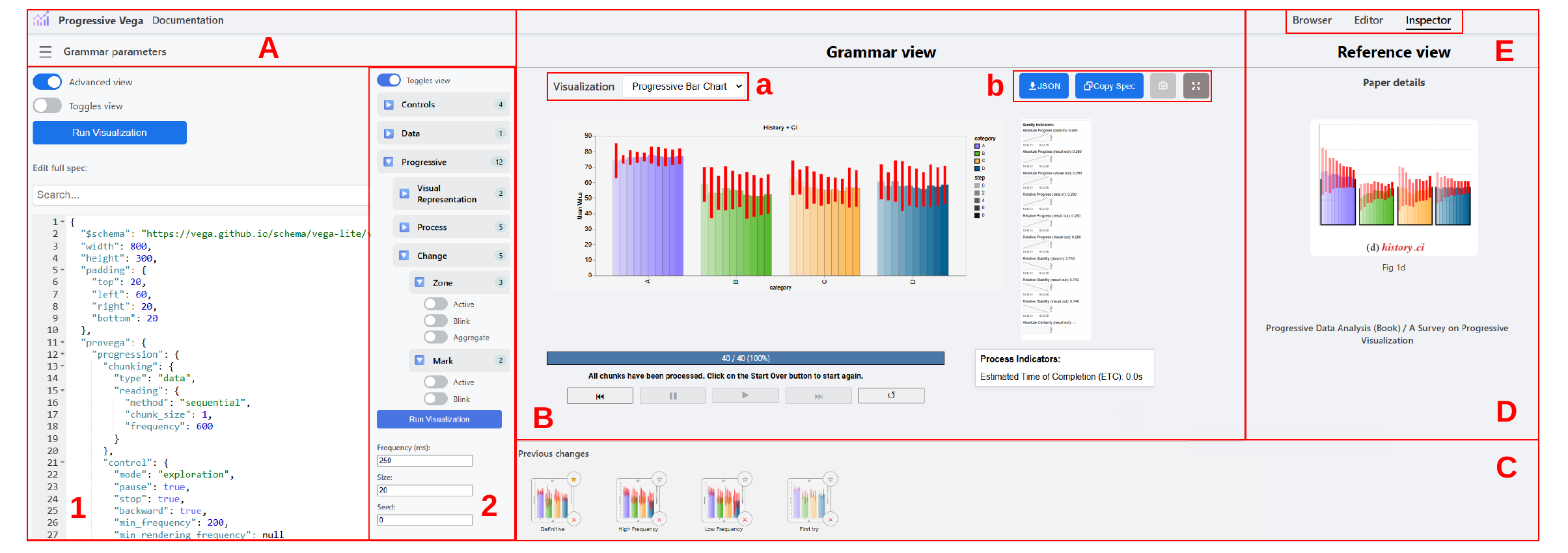}
    \vspace{-2mm}
    \caption{The Inspector View. It contains A) The Grammar Parameters View in Advanced (1) or Toggle View (2); B) The Grammar View showing a dropdown menu with the implemented examples (a) and buttons for saving specifications (b); C) The Snapshot View; D) The Reference View.}
    \label{fig:inspector}
\end{figure*}

\mypar{The Inspector View}
The inspector is Pro-Ex's main view. It allows exploration of already-implemented progressive visualizations employing the grammar, along with their equivalent representations in the paper from which they are extracted. Moreover, it allows editing the visualization's parameters and saving any resulting, unique specification in a section as a snapshot, allowing the user to quickly retrieve and test different specifications. In the top-left corner, the \emph{Documentation} text opens the grammar definition, explaining all the properties of the grammar.
As shown in Figure~\ref{fig:inspector}, it contains:


\par \noindent \textbf{A)} The Grammar Parameters View, which can be toggled/untoggled with the navigation menu toggle. Grammar parameters can be set either by activating the \textit{Toggle View}, which displays a simplified list of toggles to activate or deactivate specific properties, or by activating the \textit{Advanced View}, which displays the full, editable progressive visualization. Several helper functions are built in to ease editing of the specification (such as a search feature and the ability to enlarge/shrink the text). We opted for these two views to enable quick grammar parameter setting for novice users or in-depth editing of the specification for expert users.

\par \noindent \textbf{B)} The Grammar View, where the progressive visualization is shown, with \textbf{a)} The dropdown menu from which to select the already implemented examples and \textbf{b)} the buttons for saving the specification (as a .json file), copying the full specification, creating a snapshot, or enlarging the Grammar View in full-screen size;
\par \noindent \textbf{C)} The Snapshot View, showing the list of the saved snapshots. Clicking a snapshot restores the saved specification, which can then be renamed, saved as a favorite, or deleted. Snapshots allow quick saving and testing of specifications, enabling the user to make multiple edits to the same progressive visualization and analyze its change history.
\par \noindent \textbf{D)} The Reference View, showing reference figures and data where available for published works to share them, and the Pro-Ex view selector (\textbf{E}).

\mypar{The Editor View}
This view allows the user to create custom progressive Vega-Lite specifications. It matches the Inspector View, but without the Reference View and the guidance provided by the implemented examples. Moreover, it contains a section in its Grammar Parameters View where the user can run the specification, upload data in JSON or CSV formats, or even load a JavaScript script to generate it.
The Editor View is intended for users who are familiar with the Inspector View and want to freely test the grammar's capabilities. 


  




\section{User Evaluation}
\label{sec:evaluation}

The evaluation aimed to achieve three goals:~\textit{(i)} evaluate \provega\ expressiveness, ~\textit{(ii)} assess the usability and efficacy of Pro-Ex,~\textit{(iii)} and its support to the creation of progressive visualizations in real conditions. 

\subsection{\provega\ Expressiveness Evaluation}
To test \provega 's capability to generate progressive visualization systems, we first administered a short, comparative questionnaire lasting approximately 10 minutes. 

\mypar{Participants} The evaluation involved 39 participants. 
At the time of the evaluation, the majority of them (24) held a Master's degree, 11 held a Bachelor's degree, 2 held a High School diploma, 1 held a Doctoral Degree, and 1 held a Technical certification. Finally, their level of expertise about visualization was medium on average, as 13 answered \textit{``Low''}, 11 \textit{``Medium''}, 8 \textit{``High''}, 5 \textit{``Very Low''}, and 2 \textit{``Very High''}.

\mypar{Method and Setup} The evaluators were first presented with an informed consent form that explained the purpose and procedure of the study. After providing consent, the users were asked to compare, on a scale from 1 (completely different) to 5 (perfectly matching), the images of the 11 implemented examples with their original images, which were extracted from existing scientific literature.
Evaluators were asked to focus on overall visual and dynamic behavior similarity rather than on exact matches of data trends, position, ordering, and the shape of marks, although the examples were intended to be as close as possible to the originals. 

\begin{figure}
    \vspace{-2mm}
    \centering
    \includegraphics[width=0.85\linewidth]{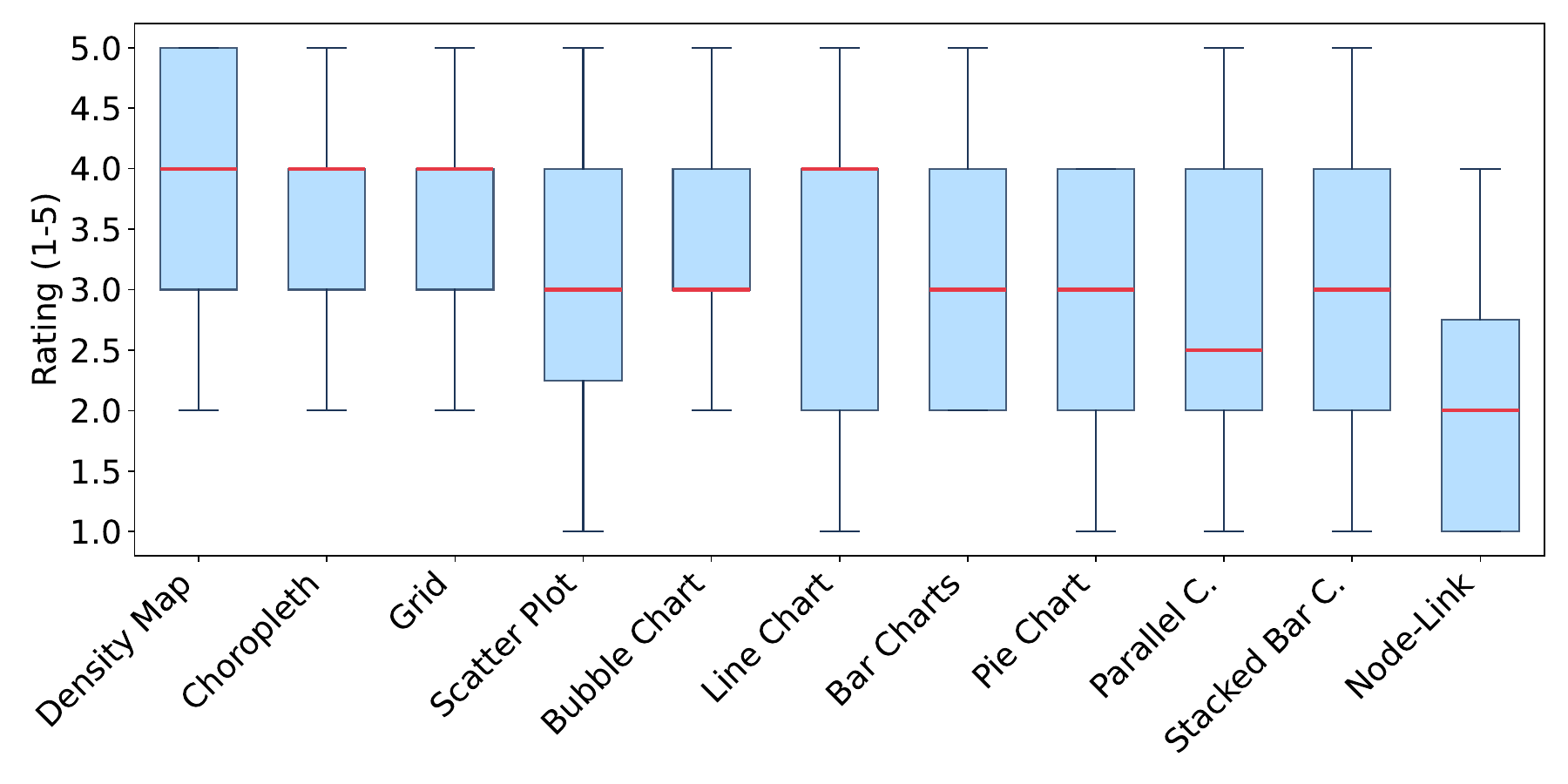}
    \vspace{-2mm}
    \caption{Similarity scores for the 11 exemplars, sorted by decreasing mean.}
    \label{fig:comparison_plot}
\end{figure}

\mypar{Results} As shown in Figure~\ref{fig:comparison_plot}, the overall mean for the scores is 3.18, while the median is 3.00.
The most similar visualizations were the Choropleth and Grid, both with an average value of 4.00. The least similar was the Node-Link visualization, which scored an average of 1.69, as it was one of the hardest to replicate exactly due to the specificity of its details. Open comments regarding this visualizations say `` \textit{Original picture has many extra labels and curved connecting edges}'', ``\textit{They seem to be two different graph knowledge representation}'', and even positive comments, telling ``\textit{The grammar generated visualization is clearer}'' and ``\textit{Different, but I’m more used to the grammar generated one (left)}''.
Several comments were biased toward exact matches in data trends, positions, order, and the mark shapes, although we reminded the evaluators at the top of each page to focus more on overall visual similarity. For instance, one of the comments for the Choropleth visualization said ``\textit{The Alaska state is smaller on the left side. Florida's color seems to be lighter on the grammar-generated visualization (left side)}''.
Overall, the \provega\ grammar was capable of generating progressive visualizations whose appearance is similar to the ones we tried to replicate, supporting goal \textit{(i)}.

\subsection{Formative Expert Evaluation}

This evaluation was structured into four tasks of increasing difficulty, each covering all possible uses of Pro-Ex and \provega. 
The tasks listed a series of steps to follow, and at the end of each step, several questions were asked, each with a score ranging from 1 to 4 (Strongly Disagree, Disagree, Agree, Strongly Agree). 
In particular, the last two tasks focused on creating a \provega\ specification using a custom dataset and then a provided one. At the end of the evaluation, participants filled out a System Usability Scale (SUS) questionnaire~\cite{sus}. Considering the high complexity of the evaluation, as evidenced by the time required to create the two custom progressive visualizations, we opted for a longitudinal study~\cite{kjaerup2021longitudinal}. The participants had 1 month in total to complete the evaluation, using a detailed Google Forms page that saved the state of the study at any moment. In this way, they could use the evaluation time span to create custom progressive visualizations for their datasets, to the best of their capabilities, reflecting real usage conditions.

\mypar{Participants} Three evaluators were involved, each with a different background and expertise in data visualization. For simplicity, we abbreviate the evaluators as “EvX”, where X = evaluator ID (i.e., Ev1, Ev2, Ev3). All of them ranged from 25 to 34 years old; Ev1 held a Master’s degree, Ev2 a Doctoral degree, and Ev3 a Bachelor’s degree. 


\begin{figure*}[t]
    \vspace{-2mm}
    \centering
    \includegraphics[width=0.9\linewidth]{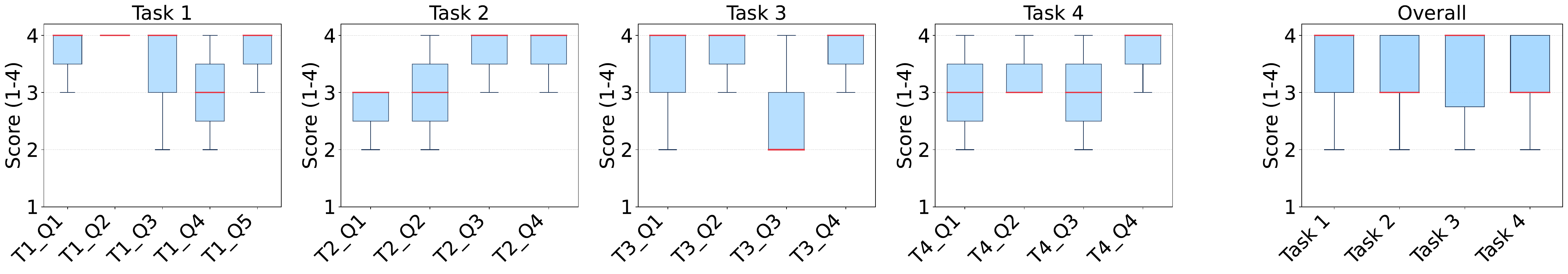}
    \vspace{-2mm}
    \caption{The evaluators' overall satisfaction of Pro-Ex for each task, sorted by decreasing median. \textbf{Task 1}: Gallery Exploration Grammar Parameters Setting;
    \textbf{Task 2}: Switching to Advanced View;
   \textbf{Task 3}: Building a custom visualization using the progressive grammar; \textbf{Task 4}: Creating a custom visualization using the progressive grammar on the FARS dataset.}
    \label{fig:satisfaction}
\end{figure*}

\subsubsection{Method and Setup}
The evaluators were first presented with an informed consent form that explained the purpose and the methodology of the longitudinal study in detail. After collecting their consent, the evaluators were shown a small tutorial to help them familiarize themselves with Pro-Ex. 
Throughout the evaluation, screenshots highlighting main components were always present to guide the evaluators. Finally, the custom grammar's documentation was always available to them.
The sequence of tasks was the following:


\mypar{(Task 1) Gallery Exploration \& Grammar Parameters Setting }
This task consisted of the Gallery View exploration and the setting of certain grammar parameters using the simplified Toggle View. Users had to select the scatterplot example, see it progress as the data chunks arrived until the very end, and change simple properties such as \textit{frequency} and \textit{seed}. They also had to activate the \textit{change} property for the marks, so the scatterplot points blinked yellow each time a new data chunk arrived (and then switched back to their intended color). Finally, the evaluators could see the results of the edits to the specifications using the snapshot functionality (Figure~\ref{fig:inspector}C).

\mypar{(Task 2) Switching to Advanced View}
Once evaluators were familiar with the Inspector View, they had to switch to the Advanced View (as shown in Figure~\ref{fig:inspector}D) and then choose the parallel coordinates visualization from the dropdown menu. The aim was to make them change the custom grammar's properties by directly modifying the specification. Finally, they activated the quality panel and carefully examined the effects of inserting and removing chunks, which could be manually controlled by activating the \textit{exploration} mode in the specification and using the progress buttons (\textit{play}, \textit{pause}, \textit{forward}, \textit{backward}).

\mypar{(Task 3) Building a custom visualization using Pro-Ex}
This task aimed to test users' capability to create a new, custom progressive visualization, using Pro-Ex. With respect to the other tasks, this one was free: evaluators had to create their own visualization using a dataset of their choice and upload it via the provided functionality. 
At least three properties of the custom grammar had to be used. Once done, all the evaluators saved and sent copies of their custom specifications and work history in Pro-Ex.

\mypar{(Task 4) Creating a custom visualization using Pro-Ex on the FARS dataset}
The final task was identical to task 3, but this time the dataset was provided by us. The evaluators had to save and send a copy of the resulting specification and history of work in Pro-Ex.

\begin{figure}[b]
    \vspace{-2mm}
    \centering
    \includegraphics[width=0.8\linewidth]{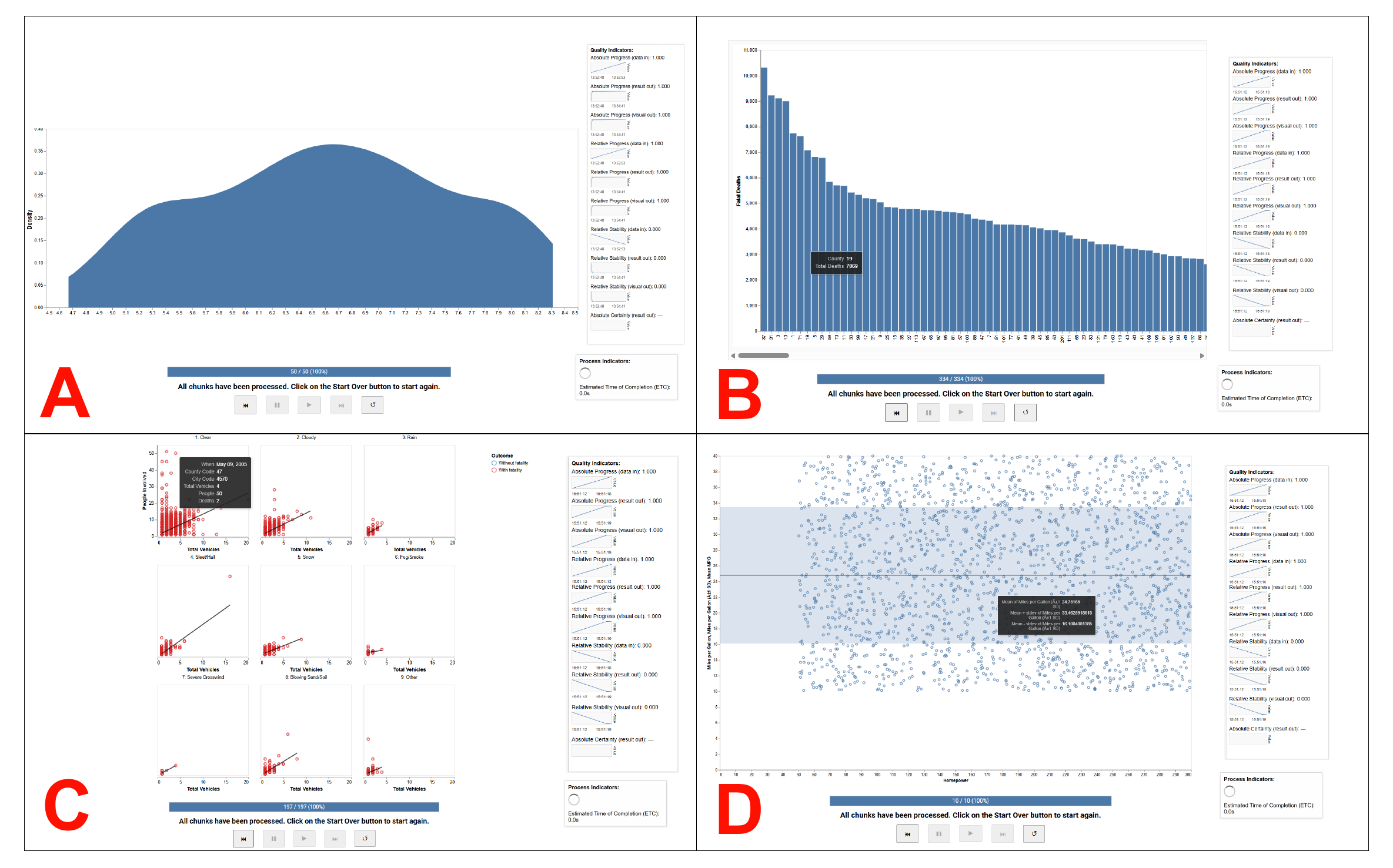}
    \vspace{-2mm}
    \caption{A selection of the custom progressive visualizations created for Task 3 (A, B) and 4 (C, D). A) univariate kernel density plot; B) a vertical bar chart (sum of fatalities by county); C) faceted scatter plot (VE\_TOTAL vs PERSONS) with an overlaid linear regression line, employing small multiples by weather code; D) scatter plot (Horsepower vs MPG) with a standard-deviation error band and a mean reference rule.}
    \label{fig:custom}
\end{figure}

\noindent Some of the created visualizations are visible in Fig.~\ref{fig:custom}.


\subsubsection{Results}

The Evaluation proved insightful in collecting feedback on Pro-Ex's capability to support users in creating custom progressive visualizations. Figure~\ref{fig:satisfaction} reports a visualization of how the evaluators felt supported in each question for each task.
Task 1, regarding the Gallery View overview and the first Inspector View exploration, had a mean score of $ 3.53$ and a median score of $4.0$. An open comment by Ev2 for the task reported ``\textit{A very short description of frequency, size, seed would be beneficial for better understanding what I am varying}'', hinting to inform the user about the grammar properties they are about to change (i.e., with a tooltip near the property name). Both Ev2 and Ev3 agreed that the quality panels were too small and on other minor UX aspects. The Advanced View exploration (Task 2) ( $\mu = 3.25$, $\tilde{\mu} = 3.0$) had a critical insight coming from Ev2. They reported, ``\textit{The advanced view seems redundant to me. In fact, most of the changes can be done through buttons more intuitively and faster. Additionally, I do not see a so fine-grained possibility of changes, as most of them can be done through buttons. Probably, the best scenario to use the advanced view is when I need to change data without reloading the entire dataset}''.
The comment hinted at giving priority to the further enhancement of the Toggle View mode, which is easier for users to understand and use at first glance, while leaving the Advanced View for other types of scenarios (i.e., modifying parameters that are not directly tied to our custom grammar, such as mark type, etc.). Ev3 highlighted other UX issues, such as the need to rerun the visualization when a user presses the ``Enter'' button while modifying the specification in the Advanced View, without having to click ``Run Visualization'' again. Tasks 1 and 2 proved successful in achieving goal \textit{(ii)}, with many insightful open comments we addressed in the current release of Pro-Ex.
Task 3, referring to the creation of a custom progressive visualization leveraging the custom grammar and the Editor View, achieved $\mu = 3.33$ and $\tilde{\mu} = 4.0$. Again, Ev2 highlighted several crucial aspects: ``\textit{When loading data, it would be useful showing what are the attributes that can be visualized''.}
Another comment reported ``\textit{The system overall looks nice, and it is intuitive to use (with small minor changes to improve usability, although it may not be the core of this work).}'' 
Finally, Ev2 stated that Pro-Ex fully supported them in building the custom progressive visualization, noting the need to give users more hints (i.e., about the grammar's properties) to avoid disrupting their workflow and to reduce the need to read the grammar's documentation many times. Figure~\ref{fig:custom} reports some of the created progressive visualizations (full set available in the supplemental material). Overall, the task scores proved to support goal \textit{(iii)}, allowing users to create their own progressive visualizations.
Task 4, tied again to the realization of a custom progressive visualization using the FARS dataset, had a $\mu = 3.25$  and a $\tilde{\mu} = 4.0$ scores. The resulting visualizations looked and worked as expected, while Ev3 suggested enhancing the load script and .csv functionality, and also reminded us to show more user feedback regarding potential errors in the specification, as they reported that sometimes the progressive visualization ran successfully but did not properly visualize the data.
The task supported the creation of custom progressive visualizations leveraging a fixed dataset, completing goal \textit{(iii)}.
\mypar{Pro-Ex Usability Evaluation} The SUS Questionnaire reported, for each evaluator, scores of 80, 90, and 40, respectively ($\tilde{\mu} = 60$). While more expert users provided a high usability score, Ev3 reported a comparatively low one.
Ev3 gave the most critical scores when answering specific questions in the questionnaire. In question 4, ``I think that I would need the support of a technical person to be able to use this system'', he answered ``Strongly Agree''; for question 7, ``I would imagine that most people would learn to use this system very quickly'', ``Strongly Disagree''; finally, for question 10, ``I need to learn a lot of things before I could get going with this system'', ``Strongly Agree''. Unfamiliarity with the Vega-Lite grammar may be a factor to consider.
To address this issue, we aim to add more user support to help users understand the grammar's properties and to provide more detailed feedback in the Editor View.


Overall, Pro-Ex was capable of supporting the evaluators in creating their custom progressive visualizations, with a fixed or custom dataset, despite minor usability issues and the need to add more support for less experienced users (e.g., easing access to \provega\ documentation for each property in the Toggle View).
\section{Discussion}
\label{sec:discussion}

In this section, we discuss the limitations and opportunities arising from the proposed grammar and environment.

\noindent\textbf{Relation with Vega-lite} As \provega\ is directly derived from Vega-lite, it inherits both pros and cons from this parent grammar. 
While we explicitly addressed scalability considerations from both visual (allowing the use of Canvas and server-side visual computation) and computational aspects (opening the use of external dedicated modules through hooks), some limitations may arise from how Vega-lite manages user interaction and the coordination of multiple views.
On the contrary, the seamless retrocompatibility with classic Vega-lite and the alignment of the grammar support the wider adoption we target for this grammar, in contrast to specific solutions.

\noindent\textbf{Relation to other progressive frameworks}
\provega\ can be easily implemented in existing progressive frameworks like ProgressiVis~\cite{progressivis} or P5 \cite{Li2020P5}. We identified two modalities for this: (i) giving control of the full progressive workflow to \provega\ and linking the specific module developed in other frameworks via hooks (see use case 4 for one example). (ii) develop the progressive pipeline in the chosen progressive framework and leverage \provega\ for the visualization aspects and auxiliary quality management (essentially like a specialized front-end component). While we recommend the first option, \provega\ can manage both.

\noindent\textbf{Leveraging Generative AI for PDAV.} a zero-shot experiment with GPT 5.2 Codex~\cite{jiang2025survey,li2022competition} showed that models can generate \provega\ specifications from documentation alone. The agent successfully implemented chunking, monitoring, and quality features, including mark highlighting. Though common generative errors, such as axis ordering~\cite{Podo2024TowardAS,10.1111:cgf.70137}, persisted (\cref{fig:app:genai}), the results demonstrate \provega's suitability for automated generation, suggesting future research into AI as an accelerator for both novice and expert users.


\newpage
\section{Conclusions}
\label{sec:conclusions}
This paper contributed \provega, a Vega-lite grammar to support the design, reproduction, and sharing of PDAV solutions.
By carefully collecting PDAV requirements from the literature and mapping them to properties and hooks, \provega\ allows both novice and expert users to implement progressive workflows.
The assisted environment Pro-Ex complements \provega, providing a platform for retrieving, sharing, and receiving assistance during prototyping and teaching.
As future work, we envision integrating Generative AI to further improve the level of assistance, based on the preliminary results discussed in the previous section.
Additionally, we plan to extend \provega's default behaviors to address emerging topics in PDAV, such as the use of guidance for progressiveness.



\bibliographystyle{abbrv-doi-hyperref}
\bibliography{biblio}

\clearpage
\newpage
\appendix

\section{Appendix}

\begin{figure}[t]
    \centering
    \includegraphics[width=0.75\linewidth]{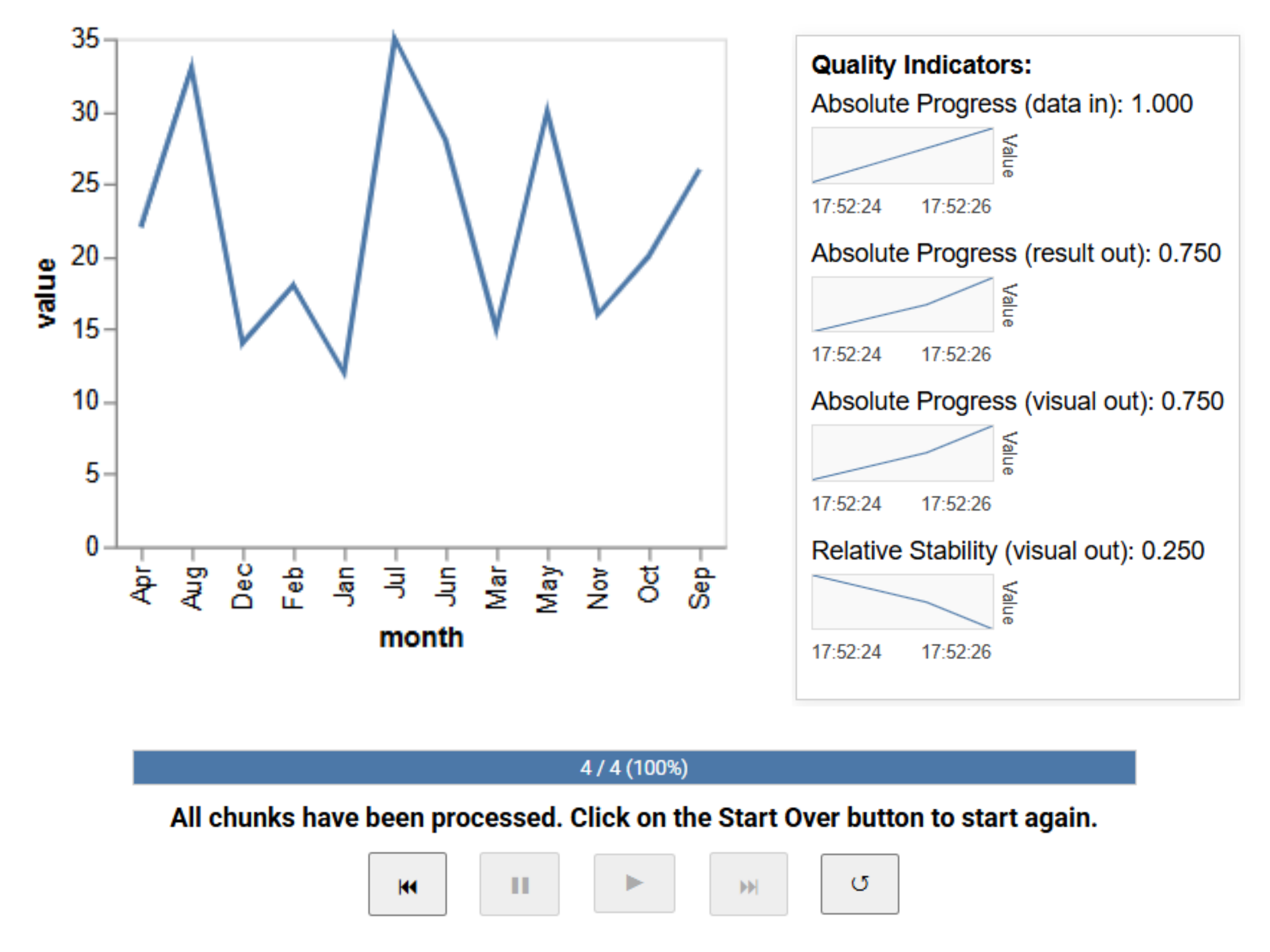}
    \caption{A simple progressive line chart obtained by passing the ProVega documentation to OpenAI's agent, GPT 5.2 Codex.}
    \label{fig:app:genai}
\end{figure}

\begin{figure*}[ht!]
\centering
\includegraphics[width=0.7\linewidth]{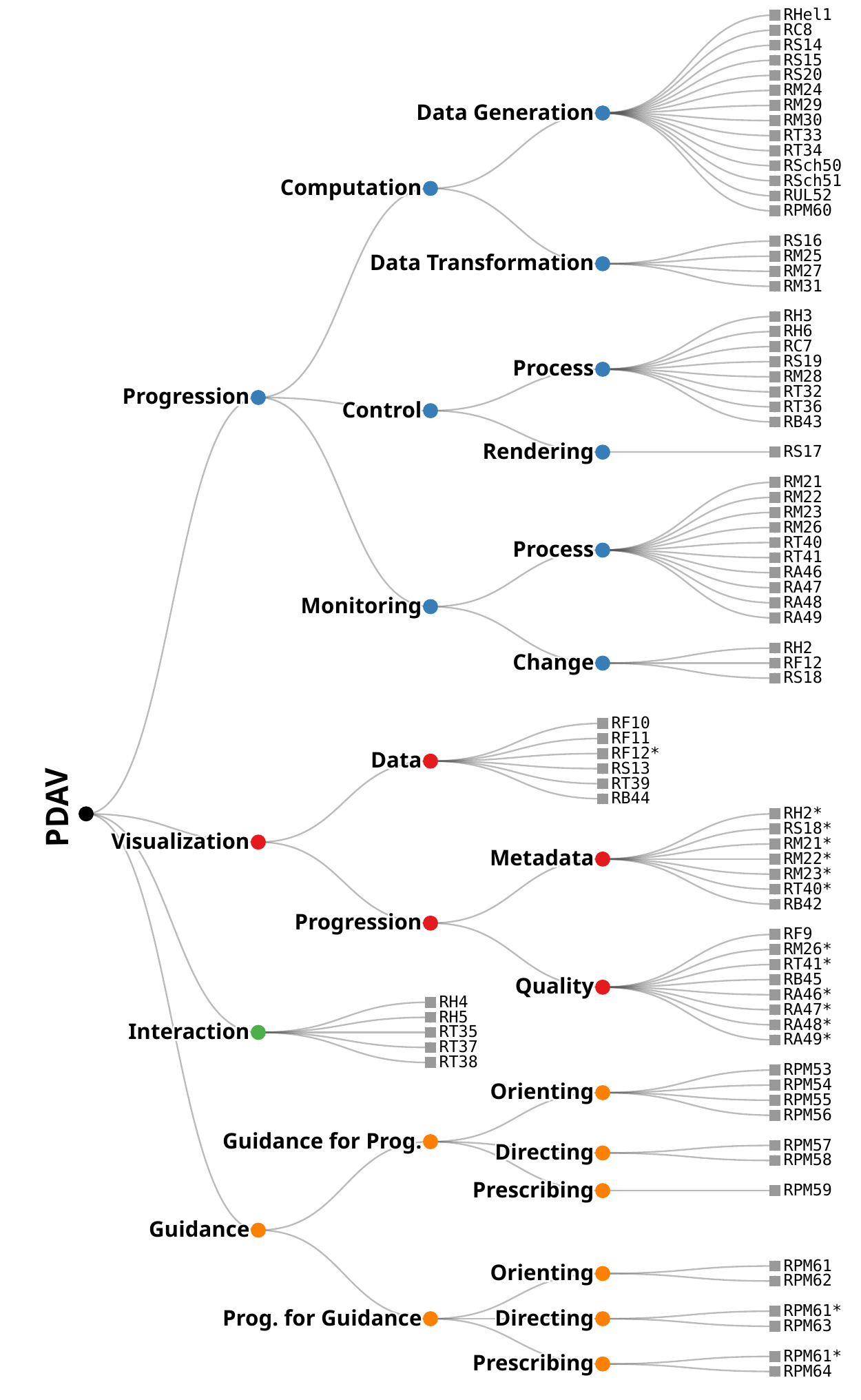}
\caption{
PDAV requirements taxonomy, serving as the foundation of the \provega\ grammar definition. The taxonomy structures PDAV requirements (\cref{sec:pva_req}) into four principal categories: {\color{taxonomyColorProgression}Progression}, {\color{taxonomyColorVisualization}Visualization}, {\color{taxonomyColorInteraction}Interaction}, and {\color{taxonomyColorGuidance}Guidance}.
Each is further subdivided into subcategories across three hierarchical levels.
Requirements may appear in multiple branches when they support more than one conceptual area; an asterisk shows such duplicated requirements. This taxonomy serves as the foundation for the \provega\ grammar definition.
}
\label{fig:app:taxonomy_big}
\end{figure*}

The appendix includes the table of characteristics of the implemented 11 PDAV exemplars (~\cref{tab:visualization_pdiv}), one example figure of the usage of Generative AI with \provega\ (~\cref{fig:app:genai}), a table with the new modeled requirements extracted from PDAV literature (~\cref{tab:app:new_pdva_req}), and a larger image of \provega\ taxonomy linked to the full set of 64 requirements from PDAV (~\cref{fig:app:taxonomy_big})

\noindent \paragraph{\textbf{Table of characteristics for the 11 implemented PDAV exemplars}} Here is the table detailing the characteristics that drove the selection of the 11 exemplars from the PDAV literature used to demonstrate the effectiveness of \provega\ in supporting the implementation of state-of-the-art progressive solutions. For all of them, the \provega\ specification is provided at \url{https://github.com/XAIber-lab/provega}. 

\begin{table}[htbp]
\small
\renewcommand{\arraystretch}{1.2} 
\centering
\caption{Classifying the reimplemented 11 exemplars from the PDAV literature by type, requirements coverage, and complexity.}
\label{tab:visualization_pdiv}

\begin{tabular}{|p{1.5cm}|p{1.9cm}|>{\centering\arraybackslash}p{1.8cm}|>{\centering\arraybackslash}p{1.2cm}|}
\hline
\textbf{Visualization \& Reference} & \textbf{Visualization Type} & \textbf{Requirements coverage} & \textbf{Complexity} \\
\hline
Pie Chart \newline {\footnotesize \cite{Sindol2012} \cite{Fekete2024}} & 
Categorical & 
\pdiv{0.5}{0.5}{0.25}{0.5} & 
Low \\
\hline
Scatter Plot \newline {\footnotesize \cite{Turkay2017} \cite{Ulmer2024}} & 
Geospatial \newline Multidimensional & 
\pdiv{0.5}{0.5}{0.5}{0.75} & 
Low \\
\hline
Grid \newline {\footnotesize \cite{Rahman2017} \cite{Ulmer2024}} & 
Temporal \newline Multidimensional \newline Color-coded & 
\pdiv{0.75}{0.75}{0.75}{0.75} & 
Low \\
\hline
History Bar Chart \newline {\footnotesize \cite{Fekete2024} \cite{Ulmer2024}} & 
Categorical \newline Layered & 
\pdiv{0.75}{0.75}{0.25}{0.75} & 
Medium \\
\hline
History Stacked Bar Chart {\footnotesize \cite{Aupetit2017} \cite{Fekete2024}} & 
Categorical \newline Layered & 
\pdiv{0.75}{0.5}{0.25}{0.75} & 
Medium \\
\hline
Parallel Coordinates {\footnotesize \cite{Rosenbaum2012} \cite{Ulmer2024}} & 
Hierarchical \newline Multidimensional & 
\pdiv{0.75}{0.5}{0.5}{0.75} & 
Medium \\
\hline
Choropleth \newline {\footnotesize \cite{Angelini2013} \cite{Ulmer2024}} & 
Geospatial \newline Color-coded & 
\pdiv{1}{0.5}{0.5}{0.75} & 
Medium \\
\hline
Line Chart \newline {\footnotesize \cite{Kesavan2020} \cite{Ulmer2024}} & 
Temporal \newline Multidimensional & 
\pdiv{0.75}{0.75}{0.75}{0.75} & 
Medium \\
\hline
Node-Link \newline {\footnotesize \cite{vanHam2009} \cite{Ulmer2024}} & 
Network & 
\pdiv{0.5}{0.25}{0.75}{0.75} & 
High \\
\hline
Bubble Chart \newline {\footnotesize \cite{Liu2018} \cite{Ulmer2024}} & 
Temporal \newline Geospatial \newline Hierarchical \newline Multidimensional & 
\pdiv{0.75}{0.75}{0.5}{1} & 
High \\
\hline
Density Map \newline {\footnotesize \cite{Schulz2016} \cite{Ulmer2024}} & 
Geospatial & 
\pdiv{1}{0.75}{1}{1} & 
High \\
\hline
\end{tabular}
\end{table}

\noindent \paragraph{\textbf{Example of experiments with Generative AI model}} In Figure~\ref{fig:app:genai} is reported an example of usage of Generative AI to generate a \provega\ specification of a progressive line chart. The prompt was \textit{``Basing on this documentation of an extension of a visualization grammar, Vega-Lite, aiming to support progressive visualizations, are you able to provide me a specification for a progressive Line chart, that necessarily uses at least three properties of the custom grammar?''}. The model was passed the documentation of \provega\ and it was capable of generating a valid specification.

\begin{table*}[!hb]
\renewcommand{\arraystretch}{1.5}
\centering
\caption{List of 19 new requirements collected from the literature on Progressive Visualization and Progressive Visual Analytics (\cref{sec:pva_req}). This list extends the set of 45 Progressive Visual Analytics requirements defined in \cite{Angelini2018Review}.}
\label{tab:app:new_pdva_req}
\resizebox{\textwidth}{!}{
\begin{tabular}{l p{.5\linewidth} l}
\toprule
\textbf{Requirement} & \textbf{Description} & \textbf{Source}\\
\midrule
\req{A46} & Provide quality indicators regarding the absolute progress & Angelini et al., 2019\cite{Angelini2019_OnQuality} (Sec. 2.3.1)\\
\req{A47} & Provide quality indicators regarding the relative progress & Angelini et al., 2019\cite{Angelini2019_OnQuality} (Sec. 2.3.1)\\
\req{A48} & Provide quality indicators regarding the relative stability & Angelini et al., 2019\cite{Angelini2019_OnQuality} (Sec. 2.3.2)\\
\req{A49} & Provide quality indicators regarding the absolute certainty & Angelini et al., 2019\cite{Angelini2019_OnQuality} (Sec. 2.3.3)\\
\midrule
\req{Sch50} & Support the Data Chunking approach for progressive processing & Schulz et al., 2016 \cite{Schulz2016Enhanced} (Sec. 2)\\
\req{Sch51} & Support the Process Chunking approach for progressive processing & Schulz et al., 2016 \cite{Schulz2016Enhanced} (Sec. 2)\\
\midrule
\req{UL52} & Support the Mixed Chunking approach for progressive processing & Ulmer et al., 2024 \cite{Ulmer2024} (Sec. 4)\\
\midrule
\req{PM53} & Provide cues about important features regarding the whole of the data or the whole progress of analysis. (G4P: Orienting) & Pérez-Messina et al., 2024 \cite{PerezMessina2024Enhancing} (Sec. 3.1.1, GT1)\\
\req{PM54} & Provide cues about data subsets, spaces, or analysis paths that are relevant for analysis. (G4P: Orienting) & Pérez-Messina et al., 2024 \cite{PerezMessina2024Enhancing} (Sec. 3.1.1, GT2)\\
\req{PM55} & Provide cues about relative interest for the task within the browsed elements. (G4P: Orienting) & Pérez-Messina et al., 2024 \cite{PerezMessina2024Enhancing} (Sec. 3.1.1, GT3)\\
\req{PM56} & Provide cues about relevant aspects of the lookup target or about related cases. (G4P: Orienting) & Pérez-Messina et al., 2024 \cite{PerezMessina2024Enhancing} (Sec. 3.1.1, GT4)\\
\req{PM57} & Provide a ranked list of actions that the user can take over the data, representation, or model (e.g., zoom into a certain area, add/delete an element, change colormap, etc.) (G4P: Directing) & Pérez-Messina et al., 2024 \cite{PerezMessina2024Enhancing} (Sec. 3.1.1, GT5-6)\\
\req{PM58} & Provide a ranked list of data cases which the user can browse. (G4P: Directing) & Pérez-Messina et al., 2024 \cite{PerezMessina2024Enhancing} (Sec. 3.1.1, GT7-8)\\
\req{PM59} & Provide a unique and complete answer to the task, either immediately encoded, step by step, or by animation. (G4P: Prescribing) & Pérez-Messina et al., 2024 \cite{PerezMessina2024Enhancing} (Sec. 3.1.1, GT9)\\
\midrule
\req{PM60} & The guidance system should incrementally generate guidance answers as partial results mature progressively. (P4G: Global) & Pérez-Messina et al., 2025 \cite{PerezMessina2025Coupling} (Sec. 4.2)\\
\req{PM61} & The guidance degree (orienting, directing, prescribing) should dynamically align with the maturity of progressive results. (P4G: Global) & Pérez-Messina et al., 2025 \cite{PerezMessina2025Coupling} (Sec. 4.2)\\
\req{PM62} & Orienting guidance should be provided from the beginning, as soon as early partial results are available. (P4G: Orienting) & Pérez-Messina et al., 2025 \cite{PerezMessina2025Coupling} (Sec. 4.3)\\
\req{PM63} & Directing guidance should be presented with mature partial results, because it reaches its ripeness only with mature partial results. (P4G: Directing) & Pérez-Messina et al., 2025 \cite{PerezMessina2025Coupling} (Sec. 4.3)\\
\req{PM64} & Prescribing guidance should be presented with definitive partial results, because it reaches its ripeness only with definitive partial results. (P4G: Prescribing) & Pérez-Messina et al., 2025 \cite{PerezMessina2025Coupling} (Sec. 4.3)\\
\bottomrule
\end{tabular}}
\end{table*}

\end{document}